%% file: Template.tex
\documentclass{article}
\usepackage{spconfa4,amsmath,graphicx}
\usepackage{IZ_Shortcuts}
\usepackage{hyperref}
\usepackage{subcaption}
\usepackage{algorithm}
\usepackage{algorithmic}
\usepackage{acronym}
\usepackage{textcomp}
\usepackage{xcolor}
\usepackage{graphicx}
\usepackage{cite}
\usepackage{comment}
\usepackage{balance}
\usepackage{acronym}
\usepackage{tabularx}
\usepackage{booktabs}
\usepackage{enumitem,bbm}
\usepackage{wrapfig}
\usepackage{hyperref}
\usepackage{soul}
\usepackage{float}
\usepackage{adjustbox}
\usepackage{tikz}

\usepackage{placeins}
\usetikzlibrary{arrows.meta,positioning,calc,shadows.blur,decorations.pathreplacing}
\usetikzlibrary{positioning,calc,decorations.pathreplacing,shadows.blur}
\definecolor{leftcol}{RGB}{255,202,128}
\definecolor{rightcol}{RGB}{173,216,230}
\definecolor{stftcol}{RGB}{210,210,255}
\definecolor{istftcol}{RGB}{210,255,210}
\definecolor{specbg}{RGB}{245,245,245}

\title{ Linearly Constrained Deep Beamformer for Multi-Speaker Scenarios}
\name{
Ilai Zaidel$^{*}$ \qquad
Ori Engel$^{*}$ \qquad
Bar Engel$^{*}$ \qquad
Sharon Gannot\qquad
\thanks{
This work was supported by the Israel Science Foundation (ISF) and the German Research Foundation (DFG) through the ISF–DFG Joint Research Program, Grant No.~1280/25. 
\\
$^{*}$These authors contributed equally to this work. }
}
\address{Faculty of Engineering, Bar-Ilan University, Israel}

\begin{document}
\ninept
\maketitle
\begin{abstract}

% We propose a deep beamforming framework for enhancing the target speaker(s) in multi-speaker environments. A deep neural network (DNN) is trained to estimate beamforming weights directly from noisy multichannel inputs, while satisfying linear spatial constraints through an adaptive multi-term loss, inspired by the augmented Lagrangian framework. The loss combines signal reconstruction with penalties that enforce a distortionless response to the target and suppress the interference subspace. The model is further guided by the target's relative transfer function (RTF) and an estimated interference subspace.
% The model can direct a beam toward the target speaker while producing nulls toward the interfering sources, while outperforming a classical LCMV beamformer constructed from
% estimated spatial signatures in several evaluated scenarios.
% Furthermore, compared with the LCMV beamformer, the proposed model produces more controlled side lobes and improved background-noise attenuation.

We propose a deep beamforming framework for enhancing target
speaker(s) in multi-speaker environments. A deep neural network
(DNN) is trained to estimate beamforming weights directly from
noisy multichannel inputs while satisfying linear spatial constraints
% through an adaptive multi-term loss inspired by the augmented
% Lagrangian framework. 
through an adaptive multi-term loss with progressively increasing constraint weights.
The loss combines signal reconstruction
with penalties that enforce a distortionless response toward the
target and suppress the interference subspace. The model is further
guided by the target relative transfer function (RTF) and the
estimated interference subspace. The proposed model can direct a
beam toward the target speaker while directing nulls toward the
interfering sources, achieving superior overall enhancement performance compared with
the classical LCMV beamformer constructed by the same estimated spatial
signatures. Furthermore, compared with the LCMV beamformer, the
proposed model produces more controlled sidelobes and improved
background-noise attenuation.

\end{abstract}
\begin{keywords}
Speech enhancement, multi-speaker, LCMV, constrained optimization, RTF estimation
\end{keywords}
\section{Introduction}
\label{sec:intro}

Multichannel beamforming enables spatial filtering of concurrent speakers using microphone arrays and is widely used for speech enhancement in complex acoustic environments. In multi-speaker scenarios, the challenge is not only to enhance the desired speaker but also to suppress interfering sources using directional filtering and null steering. In this work, we propose a fully DNN-based beamforming framework that combines deep neural networks with linearly constrained spatial objectives for multi-speaker enhancement.

% Linearly constrained beamforming provides a principled framework for controlling the spatial response of microphone arrays through explicit constraints on the beamformer weights.
Linearly constrained beamforming provides a framework for controlling microphone-array spatial responses through explicit constraints on the beamformer weights. A widely used formulation is the minimum variance distortionless response (MVDR) beamformer \cite{capon1969high,gannot2001RTF,Gannot:Consolidated_Prespective}, which preserves the desired speaker while minimizing the output noise and interference power. A more general formulation is the linearly constrained minimum variance (LCMV) beamformer \cite{Er_Cantoni_LCMV_1983}, which incorporates multiple linear constraints to explicitly shape the spatial response, enabling preservation of one or more target signals while placing nulls toward interfering sources \cite{MarkovichGannotCohen_LCMV_2009,Schwartz_Gannot_LCMV,MarkovichGannotKellermann_LCMV_2017}.

These formulations rely on accurate estimates of the sources' spatial signatures to define the beamforming constraints. Spatial information is commonly represented by the relative transfer function (RTF), which characterizes both direct-path and reverberant propagation effects. Numerous studies have demonstrated that RTF-based beamforming outperforms approaches based solely on direct-path models in terms of speech quality \cite{gannot2001RTF,Shmaryahu_Gannot_acoustic_reflections_2022}.
 LCMV beamforming can achieve perfect interference cancellation when accurate estimates of the target and interfering speakers' RTFs (or basis vectors spanning their respective subspaces) are available. This motivates the development of robust RTF estimation methods, e.g., \cite{Markovich:Covariance_Whitening_Analysis}.

Recently, DNN-based beamformers have achieved strong performance by jointly learning spatial and spectral representations from data \cite{FasNET,Ren:Causal_UNet_Beamforming}. However, these methods are often not directly interpretable, motivating approaches that explicitly analyze and encourage spatial selectivity in multichannel processing \cite{Kellermann_Spatial__info_TSE,tesch2022insights,Cohen:ExNet_BF_PF,Habets_Beampatterns}. Prior works have incorporated RTF information either by estimating classical beamformer coefficients \cite{Gannot:Consolidated_Prespective,ronai2025rtf,RTF_estimation_correlations} or by embedding it as a spatial filter within deep architectures \cite{Align_and_Filter}. Building on these advances, our previous work \cite{zaidel2026binauraldeepbeamforming} demonstrated that guiding a DNN-based beamformer with time-varying RTF estimates improves the spatial consistency and directional behavior of learned beamformers. %Nevertheless, such approaches  do not explicitly enforce any directional constraints or interference nulls.
% Nevertheless, such approaches do not explicitly enforce spatial constraints and therefore cannot reliably guarantee directional null steering or distortionless target preservation.
% Nevertheless, although some approaches incorporate spatial guidance or estimate interference-related quantities required for constrained beamforming, they do not directly train the beamforming weights themselves under explicit linear spatial constraints.
Nevertheless, such approaches only enforce spatial guidance toward the desired speaker, without explicitly constraining the response toward interfering sources, and therefore cannot reliably guarantee directional null steering.

Several works have explored incorporating directional constraints and null steering into learning-based beamforming frameworks. Works such as \cite{chazan2018DNN_LCMV,yang2024interference} combine deep learning with classical constrained beamforming formulations, where neural networks estimate spatial or statistical quantities required for beamforming and interference representations, while the final spatial filtering is still performed using analytical model-based beamformers. In parallel, works such as \cite{Karimi_Nulling_Control}, proposed in the context of wireless communications, train DNNs to directly predict beamforming weights that imitate the behavior of constrained beamformers and achieve directional null control. However, in such approaches, spatial constraints are not explicitly incorporated into the learning objective; instead, they are either enforced analytically via a separate beamforming stage or approximated implicitly through supervised training.

Our work builds upon the frameworks of \cite{Cohen:ExNet_BF_PF,zaidel2026binauraldeepbeamforming}, which learn beamforming weights using a DNN while preserving the beamforming structure. While both works demonstrate strong enhancement capabilities, including dynamic binaural beamforming in \cite{zaidel2026binauraldeepbeamforming}, they do not explicitly enforce spatial constraints and therefore cannot reliably achieve directional null steering toward interfering sources.

In contrast, we propose a fully DNN-based beamformer optimized to satisfy linear constraints for target speaker extraction and interference suppression. Specifically, we design an adaptive multi-term loss function, reminiscent of the LCMV criterion, that combines signal reconstruction with constraint-driven penalties that enforce a distortionless response toward the target and null out the interference subspace.
% The training process follows an augmented Lagrangian-inspired approach, where the weights of the constraint terms are gradually increased during training, enabling the network to learn both accurate signal reconstruction and spatially selective filtering.
The weights of the constraint terms are gradually increased during training, enabling the network to learn both accurate signal reconstruction and spatially selective filtering.
To further guide the learning process, the model is provided with the estimated target RTF and basis vectors spanning the interference subspace, which together define the constraint structure and promote spatially selective beamforming behavior.

% Throughout this work, we evaluate the learned beamformer under three configurations: (i) guidance using estimated target RTFs and interference subspaces (referred to as the ``Estimated RTF'' model), (ii) a model without RTF guidance (``No RTF''), and (iii) guidance using the oracle RTFs of the target and interfering speakers (``Oracle RTF'').
Throughout this work, we evaluate the learned beamformer under three configurations:
(i) guidance using the estimated target RTF and interference subspace
(``Estimated RTF''),
(ii) no RTF guidance (``No RTF''), in which the RTF
inputs are replaced with zeros while the network
architecture remains unchanged, and
(iii) guidance using the oracle target RTF and interference subspace
(``Oracle RTF'').

\section{Problem Formulation}
\label{sec:format}

In the short-time Fourier transform (STFT) domain, the multichannel mixture signal is modeled as
\begin{equation}\label{eq:stft_model}
\vct{y}(l,k)=\vct{H}(k)\vct{s}(l,k)+\vct{n}(l,k) \in \mathbb{C}^{M\times 1},
\end{equation}
where $l=0,\ldots,L-1$ and $k=0,\ldots,F-1$ denote the time-frame and frequency-bin indices, respectively, and $M$ is the number of microphones. Here,
\begin{equation}
\vct{s}(l,k)=
\begin{bmatrix}
s_1(l,k), \dots, s_J(l,k)
\end{bmatrix}^{\top},
\end{equation}
represents the $J\leq M$ active speakers, and
\begin{equation}
\vct{H}(k)=
\begin{bmatrix}
\vct{h}_1(k), \dots, \vct{h}_J(k)
\end{bmatrix}
\end{equation}
comprises the acoustic transfer functions (ATFs) from each source to the microphones. 
The vector $\vct{n}(l,k)$ denotes additive noise.
We apply the time-invariant spatial filter
\begin{equation}\label{eq:enhancedsignal}
\hat{s}(l,k)=\mathbf{w}^{\mathrm{H}}(k)\vct{y}(l,k),
\end{equation}
where $\mathbf{w}(k)$ denotes the DNN based beamformer weights and $\hat{s}(l,k)$ is the beamformer output.
The output is designed to estimate, through the optimization of a suitable loss function, a target signal defined as a linear combination of the sources of interest:
\begin{equation}\label{target}
s_{\rm target}(l,k)=\vct{g}^{\top}\vct{s}(l,k),
\end{equation}
where $\vct{g}\in \mathbb{R}^{J \times 1}$ is a weighting vector.
Typically, the entries of $\vct{g}$ are `1' for desired source(s) and `0' for interference sources.
In the proposed method, the beamformer weights are chosen to minimize the loss between $\hat{s}$ and $s_{\rm target}$, while satisfying linear constraints.

\section{Proposed Method}
\label{sec:pagestyle}

This section presents the proposed DNN-based beamforming framework. The model builds on the U-Net architecture of \cite{Cohen:ExNet_BF_PF,zaidel2026binauraldeepbeamforming}. We introduce the proposed spatial guidance based on estimates of the target speaker's RTF and the interference subspace, together with a new loss function for improved beampattern control. The complete architecture is illustrated in Fig.~\ref{fig:proposed model}.

\begin{figure*}[t]
%\vspace*{-4mm} % adjust to pull figure up

\centering
\begin{adjustbox}{max width=\textwidth}
\input{figures/proposed_model}
\end{adjustbox}
\setlength{\belowcaptionskip}{-12pt}
\setlength{\abovecaptionskip}{-10pt}
\caption{Overview of the proposed DNN-based beamforming.}
\label{fig:proposed model}
%\vspace*{-2mm}
\end{figure*}

\subsection{Attention Fusion for RTF Guidance}

The frontend of the U-Net beamformer employs an attention mechanism to fuse the estimated target RTF and interference subspace with the multichannel microphone signals, with dimensions $\mathbb{C}^{M\times F\times L}$. Each spatial cue (the target RTF or an interference basis vector), of size $\mathbb{C}^{M\times F}$, is broadcast along the temporal dimension to match the input size. Each of the $J$ spatial cues is separately processed by an identical cross-attention module with shared weights. To reduce the computational and memory cost of full temporal attention, cross-attention is applied independently at each frequency bin over non-overlapping windows of 24 frames, where the microphone signal features serve as queries and the spatial cue features as keys and values. For each spatial cue, the shared module produces a fused feature map with the same time-frequency resolution as the input. The resulting $J$ fused feature maps are concatenated with the microphone signal features and fed to the U-Net encoder. 
% Finally, a fully connected layer predicts the complex beamforming weights, followed by complex normalization and a learnable global gain.

\subsection{RTF Estimation}

% To estimate the static spatial signatures of the speakers, we employ the covariance-whitening (CW) method for RTF estimation  \cite{Markovich:Covariance_Whitening_Analysis}. Given $J$ active speakers, our goal is to preserve the target speaker while suppressing the remaining $J-1$ interfering speakers. To this end, we apply the same CW procedure to different speaker-partitioned frame sets: target-only segments are used to estimate the target RTF, whereas interference-only segments are used to estimate the subspace spanned by the interfering speakers.  In addition, noise-only segments are assumed to be available for estimating the noise covariance matrix.

% To estimate the static spatial signatures of the speakers, we employ the covariance-whitening (CW) method for RTF estimation \cite{Markovich:Covariance_Whitening_Analysis}. Although the formulation in \eqref{target} is general, here we consider single-target extraction. Given $J$ active speakers, our goal is therefore to preserve one target speaker while suppressing the remaining $J-1$ interfering speakers. To this end, we apply the same CW procedure to different speaker-partitioned frame sets: target-only segments are used to estimate the target RTF, whereas interference-only segments are used to estimate the subspace spanned by the interfering speakers. In addition, noise-only segments are assumed to be available for estimating the noise covariance matrix.

To estimate the speakers' spatial signatures in a static acoustic environment, we employ the covariance whitening (CW) method for RTF estimation~\cite{Markovich:Covariance_Whitening_Analysis}. Although the formulation in \eqref{target} is general, this work focuses on single-target extraction. Given $J$ active speakers, the objective is to preserve the target speaker while suppressing the remaining $J-1$ interferers.
To this end, the same CW procedure is applied to frame sets corresponding to different source activity patterns, assuming such annotations are available. Frames containing only the target speaker are used to estimate the target RTF, whereas frames containing only interfering speakers are used to estimate the interference subspace. In addition, noise-only frames are assumed to be available for estimating the noise covariance matrix.
\subsubsection{Covariance-Whitening}

Let $\mathcal{V}_n$ denote the set of frames in which only noise is present. 
The noise covariance matrix is then estimated as
\begin{equation}\label{eq: Noise Correlation Estimation_short}
\hat{\Mat{\Phi}}_{\vct{nn}}(k)
=
\frac{1}{|\mathcal{V}_n|}
\sum_{l \in \mathcal{V}_n}
\vct{y}(l,k)\vct{y}^\rmH(l,k).
\end{equation}
The whitening operation is defined as:
\begin{equation}\label{eq: Whitened Measurement_short}
\vct{y_{w}}(l,k)
=
\hat{\Mat{\Phi}}^{-1/2}_{\vct{nn}}(k)\,\vct{y}(l,k),
\end{equation}
where $\hat{\Mat{\Phi}}^{-1/2}_{\vct{nn}}(k)$ is computed via eigenvalue decomposition (EVD) of $\hat{\Mat{\Phi}}_{\vct{nn}}(k)$.

Let $\mathcal{V}_t$ denote the set of frames in which only the target speaker is active, and let $\mathcal{V}_i$ denote the set of frames in which only interfering speakers are active (in both cases, the stationary noise is also present). The set $\mathcal{V}_i$ includes multiple active sources of the interference group, but not the target source.
For a given frame set $\mathcal{V}$, the corresponding noisy covariance matrix is estimated as:
\begin{equation}
\hat{\Mat{\Phi}}_{\vct{yy}}^{(\mathcal{V})}(k)
=
\frac{1}{|\mathcal{V}|}
\sum_{l \in \mathcal{V}}
\vct{y}(l,k)\vct{y}^{\rmH}(l,k),
\end{equation}
and the whitened covariance matrix is defined by
\begin{equation}
\hat{\Mat{\Phi}}_{\vct{y_w y_w}}^{(\mathcal{V})}(k)
=
\hat{\Mat{\Phi}}^{-1/2}_{\vct{nn}}(k)\,
\hat{\Mat{\Phi}}_{\vct{yy}}^{(\mathcal{V})}(k)\,
(\hat{\Mat{\Phi}}^{-1/2}_{\vct{nn}})^\rmH(k).
\end{equation}
Applying this procedure to $\mathcal{V}_t$, the target RTF is obtained from $\hgvct{\psi}^{(t)}$ the dominant eigenvector of $\hat{\Mat{\Phi}}_{\vct{y_w y_w}}^{(\mathcal{V}_t)}(k)$:
\begin{equation}\label{eq:RTF_target_CW}
\hvct{a}^{(t)}(k)
=
\frac{\hat{\Mat{\Phi}}^{\rmH/2}_{\vct{nn}}(k)\hgvct{\psi}^{(t)}}
{\vct{e}^{\top}_{\mathrm{ref}}\hat{\Mat{\Phi}}^{\rmH/2}_{\vct{nn}}(k)\hgvct{\psi}^{(t)}},
\end{equation}
with $\vct{e}_{{\mathrm{ref}}}$ the selection vector for the chosen reference microphone. 

Applying the same procedure to $\mathcal{V}_i$, the interference subspace is estimated by taking the $J-1$ dominant eigenvectors of $\hat{\Mat{\Phi}}_{\vct{y_w y_w}}^{(\mathcal{V}_i)}(k)$. Denoting these eigenvectors by $\{\hgvct{\psi}^{(i)}_{j}\}_{j=1}^{J-1}$, the corresponding interference basis vectors are obtained:
\begin{equation}
\label{eq:RTF_int_CW}
\hvct{u}^{(i)}_{j}(k)
=
\frac{\hat{\Mat{\Phi}}^{\rmH/2}_{\vct{nn}}(k)\hgvct{\psi}^{(i)}_{j}}
{\vct{e}^{\top}_{\mathrm{ref}}\hat{\Mat{\Phi}}^{\rmH/2}_{\vct{nn}}(k)\hgvct{\psi}^{(i)}_{j}},
\qquad j=1,\dots,J-1.
\end{equation}
% which are stacked into
% \begin{equation}
% \Mat{A}_{\mathrm{CW}}^{(i)}(k)
% =
% \left[
% \hvct{a}_1^{(i)}(k),\,
% \hvct{a}_2^{(i)}(k)
% \right].
% \end{equation}
%
Thus, the same CW procedure is used for both estimations: the dominant eigenvector from $\mathcal{V}_t$ provides the target RTF, while the dominant eigensubspace from $\mathcal{V}_i$ provides a basis that spans the interfering speakers' RTF subspace. 
% The application of the EVD requires $\mathcal{O}(M^3)$ operations per frequency, in addition to the computational cost of the whitening procedure.
% The estimated interference subspace is used as spatial guidance for the
% network input. In contrast, the training objective in Sec.~3.3 employs
% the ground-truth interfering-speaker RTFs, denoted by
% $\Mat{A}_{\mathrm{interf}}(k)$, to explicitly define the null-steering
% constraints during supervised training.

\subsection{Loss Function and Training Process}

% To enforce the constrained optimization, we adopt a loss function inspired by the augmented Lagrangian framework, where constraint terms are incorporated into the objective as weighted penalties, with coefficients that are iteratively rescaled during training \cite{bertsekas2014constrained}.
In this work, we propose to minimize the negative SI-SDR between the estimated source and the target combination of the sources of interest. We discuss a single desired source and $J-1$ interference sources, namely $\vct{g}^{\top}=[1,0,\ldots,0]$ (assuming, without loss of generality, that the desired source is source \#1). An extension to multiple desired sources is straightforward. 
% To impose a distortionless response to the desired source and a null response to the interference source, we define the following loss function, reminiscent of the LCMV criterion. 
To impose a distortionless response toward the target source and
suppress the interference sources, we define the following loss
function inspired by the LCMV criterion.

% We adopt a procedure inspired by the augmented Lagrangian framework~\cite{bertsekas2014constrained}, in which the constraint terms are incorporated into the objective as weighted penalties. The corresponding penalty coefficients are progressively increased during training, thereby encouraging gradual satisfaction of the spatial constraints.
The constraint terms are incorporated into the objective as weighted penalties. Their corresponding coefficients are progressively increased during training following a predefined schedule, allowing the network to first learn accurate signal reconstruction before increasingly enforcing the spatial constraints.

The network predicts frequency-dependent time-varying beamforming weights $\mathbf{w}_l(k)\in\mathbb{C}^M$, which are averaged over the time-frame index $l$ to obtain the final time-invariant beamformer weights $\mathbf{w}(k)\in\mathbb{C}^M$.  The enhanced signal $\hat{s}$ and the target signal $s_{\mathrm{target}}$, defined in \eqref{eq:enhancedsignal} and \eqref{target}, respectively, are considered here in the time domain. 
The training objective is given by:
\begin{equation}\label{the loss}
\begin{aligned}
\mathcal{L}
&=
-\mathrm{SI\text{-}SDR}(\hat{s},s_{\mathrm{target}}) \\
&\quad+
\lambda_{\mathrm{pass}}
\,\mathbb{E}_{k}\!\left[
\left|\mathbf{w}^{\rmH}(k)\mathbf{a}_{\mathrm{target}}(k)-1\right|^2
\right] \\
&\quad+
\lambda_{\mathrm{null}}\,
\mathbb{E}_{k}\!\left[
10\log_{10}\!\left(
\left\|\mathbf{w}^{\rmH}(k)\mathbf{A}_{\mathrm{interf}}(k)\right\|^2+\epsilon
\right)
\right],
\end{aligned}
\end{equation}
which jointly promotes target reconstruction, enforces a distortionless response toward the desired direction, and encourages null steering toward the interference subspace.
Here, $\mathbf{a}_{\mathrm{target}}(k)\in\mathbb{C}^{M}$ denotes the oracle RTF of the target speaker used for supervised training, and $\mathbf{A}_{\mathrm{interf}}(k)\in\mathbb{C}^{M\times(J-1)}$ contains the oracle RTFs of the interfering speakers.
Applying the null penalty on a logarithmic scale increases the sensitivity to low-level residual interference, thereby encouraging deeper nulls compared with linear-domain penalties.
The penalty weights $\lambda_{\mathrm{pass}}$ and $\lambda_{\mathrm{null}}$ are gradually increased during training according to a predefined schedule, and are activated only after an initial warm-up period of 10 epochs.
In this work, the target speaker was selected at random from the $J$ speakers.

It is important to note that the network is trained to optimize the loss function in \eqref{the loss} using oracle spatial information for supervision. However, during inference, oracle RTFs are not available. Instead, the network is guided by the estimated target RTF in \eqref{eq:RTF_target_CW} and the estimated interference subspace in \eqref{eq:RTF_int_CW}, which are provided as inputs to the network.

\section{Experimental Study}
\label{sec:Experimental Study}
This section details the dataset generation process and presents the results of the proposed model.

\subsection{Dataset Generation and Noise Environment}

Multichannel multi-speaker recordings were simulated in randomly generated acoustic environments. Each sample corresponds to a room with width and length uniformly drawn in $[6,9]$~m and a fixed height of $3$~m.
% An $8$-microphone linear array was placed at a height of $1.3$~m and randomly tilted within $[-45^\circ,45^\circ]$ (see Fig.~3 in \cite{Cohen:ExNet_BF_PF} for the array configuration). 
An $8$-microphone nonuniform linear array with microphone inter-distances (relative to the array center) of
$[-17,-12,-7,-4,4,7,12,17]$ cm was randomly positioned
% (subject to the room boundaries) 
and randomly rotated within $[-45^\circ,45^\circ]$, following~\cite{Cohen:ExNet_BF_PF}.
Speech signals were drawn from the LibriSpeech dataset \cite{panayotov2015librispeech} and positioned at a distance of $1$--$1.5$~m from the array center. Each sample includes $J\in\{2,3\}$ static speakers immersed in a stationary babble-noise environment, with one target speaker and $J-1$ interfering speakers. Both anechoic and reverberant target/interference conditions were considered. 
% The stationary babble noise was pre-generated using the room impulse response (RIR) generator \cite{habets2006room} by summing $20$ simultaneously active speakers positioned near the room walls, and was introduced to enable noise covariance estimation. To avoid the computational cost of generating the full acoustic scene using GPU-RIR, only the reverberant target and interfering speakers were simulated using the GPU-RIR package \cite{diaz2021gpurir} with $T_{60}\in[0.3,0.55]$~s, whereas the babble-noise signals and anechoic target/interfering signals were generated using the RIR generator \cite{habets2006room}. 
For each sample, stationary babble noise was generated by summing $20$ randomly selected active speakers positioned near the room walls using the room impulse response (RIR) generator \cite{habets2006room}. 
%This noise was included to enable noise covariance estimation. 
The same simulator was used to generate the anechoic target and interfering speakers. Reverberant target and interfering speakers with $T_{60}\in[0.3,0.55]$~s were simulated using the GPU-RIR package \cite{diaz2021gpurir}, while the babble noise remained anechoic to reduce computational cost.
Each recording contains a $4$~s segment used to estimate the DNN-based beamformer. The first $0.5$~s contains only babble noise, followed by a $1$~s target-only segment and a $1$~s interference-only segment. The final $1.5$~s contains the mixture with all speakers simultaneously active.

The estimated time-invariant beamforming weights are then applied to an additional $4$~s fully overlapped mixture, used exclusively for evaluation to provide a sufficiently long recording for reliable performance assessment. The training set contains $20{,}000$ multichannel recordings.

% The estimated time-invariant beamforming weights are then applied to an additional $4$~s fully overlapped mixture segment, resulting in final $8$~s recordings used for evaluation.

\subsection{Results}
This section reports the results of the proposed linearly constrained DNN beamformer. Audio samples, beampatterns, and implementation code are available in our online repository.\footnote{\texttt{https://github.com/GannotLab/LC-DeepBeam}}

\vspace{2pt}\noindent\textbf{Enhancement Performance:}
We evaluate the proposed method using SI-SDR, SNR, and SIR, computed over the active mixture frames spanning $2.5$--$8$~s of the $8$~s recordings, during which all speakers are simultaneously active. In addition, we report the power ratio $\mathrm{Pwr~Ratio}=10\log_{10}\!\left(\frac{\mathbb{E}|x_{\mathrm{out}}|^2}{\mathbb{E}|x_{\mathrm{in}}|^2}\right)$, 
% which is computed by applying the learned beamformer weights separately to each signal component and measuring its average energy before and after beamforming. All outputs are normalized to preserve the target speaker power.
computed separately for each signal component after applying the learned beamformer. The beamformer is first normalized to preserve the target speaker power, such that the target $\mathrm{Pwr~Ratio}$ is fixed at $0\,\mathrm{dB}$. The reported values therefore represent the attenuation of the interfering speakers and background noise relative to the preserved target.

As a competing method, the analytical LCMV beamformer is constructed using the estimated spatial signatures and is given by:
\begin{equation}
\mathbf{w}_{\mathrm{LCMV}}(k)
=
\hat{\mathbf{\Phi}}_{nn}^{-1}(k)\mathbf{C}(k)
\left(
\mathbf{C}^H(k)\hat{\mathbf{\Phi}}_{nn}^{-1}(k)\mathbf{C}(k)
\right)^{-1}
\mathbf{g},
\end{equation} 
where $\mathbf{C}(k)=\left[\hat{\mathbf{a}}^{(t)}(k),
\hat{\mathbf{u}}^{(i)}_1(k),
\ldots,
\hat{\mathbf{u}}^{(i)}_{J-1}(k)\right]$, $\hat{\mathbf{\Phi}}_{nn}(k)$ defined in \eqref{eq: Noise Correlation Estimation_short}, and the spatial signatures are defined in
\eqref{eq:RTF_target_CW} and \eqref{eq:RTF_int_CW} for the target source and the interference subspace, respectively.

% Tables~\ref{tab:beamforming_3spk_no_reverb} and \ref{tab:beamforming_2spk_reverb} summarize the performance of the proposed beamformer under both anechoic and reverberant conditions for two- and three-speaker mixtures ($J\in\{2,3\}$).
% Although the LCMV beamformer explicitly enforces spatial
% constraints, the proposed learned models achieve superior overall
% enhancement performance together with stronger background-noise
% attenuation. 
% In particular, the proposed models achieve substantially higher SI-SDR and SNR values than the LCMV baseline in both the anechoic and reverberant scenarios, while maintaining competitive interference suppression. While the LCMV beamformer achieves stronger suppression for some interference metrics, the proposed models consistently provide better overall enhancement performance in terms of SI-SDR, SNR, and background-noise suppression.
% The ``Estimated RTF'' and
% ``No RTF'' models achieve similar enhancement metrics in
% the evaluated scenarios.

Tables~\ref{tab:beamforming_2spk_reverb} and \ref{tab:beamforming_3spk_no_reverb} summarize the performance of the proposed beamformer under both anechoic and reverberant conditions for two- and three-speaker mixtures ($J\in\{2,3\}$). Compared with the LCMV beamformer, the proposed models consistently achieve higher SI-SDR and SNR in both anechoic and reverberant scenarios, together with stronger background-noise attenuation and competitive interference suppression. While the LCMV beamformer exhibits stronger suppression for some metrics, the proposed models provide better overall enhancement performance. The ``Estimated RTF'' and ``No RTF'' models exhibit similar performance.

\vspace{2pt}\noindent\textbf{On the Importance of RTF Guidance:}
To further examine the role of spatial guidance,
Table~\ref{tab:beamforming_3spk_no_partition} presents a
conceptual fully overlapped scenario in which all speakers remain
simultaneously active throughout the recording. Since the proposed
CW-based estimation method requires separated source activity
patterns, this experiment serves only to illustrate the importance of
spatial guidance. In this setting, the unguided model (``No RTF'')
fails to achieve meaningful enhancement or interference
suppression, whereas the ``Oracle RTF'' model maintains strong
directional filtering and null steering.
These results motivate the development of spatial-guidance methods
that do not rely on separated source activity patterns.

\vspace{2pt}\noindent\textbf{Beampattern Analysis:}
% The narrowband beampattern is defined as $B(k,\theta)=\vct{w}^{\rm H}(k)\vct{h}(k,\theta)$, where $\vct{h}(k,\theta)$ is the steering vector corresponding to the direction of arrival $\theta$. Figure~\ref{fig:beampattern_collage_3_speakers} presents the corresponding wideband beampower, computed as $P(\theta)=\sum_k |B(k,\theta)|^2$.
% % Compared with the oracle LCMV beamformer, the proposed learned models produce shallower interference nulls, but exhibit lower sidelobe levels, leading to improved background noise suppression. 
% Compared with the LCMV beamformer, the proposed learned models
% produce more spatially selective and directional responses, with
% lower sidelobe levels and improved background-noise suppression.
% While the LCMV beamformer maintains explicit null constraints,
% its spatial response is less structured and exhibits stronger sidelobes.
% In contrast to the unguided model, the RTF guided beamformers produce more directional and spatially coherent responses, with well-focused main lobes toward the target direction and clearer nulls toward the interfering speakers. The unguided model exhibits a less structured beampattern despite achieving similar enhancement metrics. These observations suggest that RTF guidance encourages the network to learn spatially selective filtering behavior, whereas the unguided model relies more heavily on spectral filtering cues.
The narrowband beampattern is defined as
$B(k,\theta)=\vct{w}^{\rm H}(k)\vct{h}(k,\theta)$, where
$\vct{h}(k,\theta)$ is the steering vector corresponding to the
direction of arrival $\theta$. Figure~\ref{fig:beampattern_collage_3_speakers}
presents the corresponding wideband beampower, computed as
$P(\theta)=\sum_k |B(k,\theta)|^2$.
Compared with LCMV, the proposed models produce more spatially selective responses, characterized by lower sidelobes and narrower main lobes toward the target. RTF guidance further yields more spatially coherent beampatterns than the unguided model, despite both achieving similar speech enhancement metrics. Although the LCMV beamformer provides stronger suppression according to some objective metrics, Fig.~\ref{fig:beampattern_collage_3_speakers} and the examples in the online repository demonstrate that the proposed beamformer often produces more spatially selective responses with lower sidelobes.
\begin{table}[htbp]
\centering
\scriptsize
\setlength{\tabcolsep}{5pt}
\renewcommand{\arraystretch}{1.25}
\caption{Two-speaker scenario (reverberant target/interference).}
\label{tab:beamforming_2spk_reverb}

\vspace{1mm}

\begin{tabular}{l|c|c|c|c|c}
\hline
\textbf{Metric [dB]} & \textbf{Input} & \textbf{Est. RTF} & \textbf{No RTF} & \textbf{Oracle RTF} & \textbf{LCMV} \\
\hline
\multicolumn{6}{c}{\textbf{Target speaker (enhancement)}} \\
\hline
SI-SDR & -1.81 & 0.33 & 0.05 & 0.40 & -3.50 \\
SNR    &  3.30 & 5.61 & 6.33 & 6.11 & 5.24 \\
SIR    & -0.03 & 4.78 & 4.62 & 5.00 & 5.58 \\
% Pwr Ratio & -- & 0.00 & 0.00 & 0.00 & 0.00 \\
\hline
\multicolumn{6}{c}{\textbf{Interferer 1 (suppression)}} \\
\hline
Pwr Ratio  & -- & -4.81 & -4.66 & -5.03 & -5.61 \\
\hline
\multicolumn{6}{c}{\textbf{Background noise (suppression)}} \\
\hline
Pwr Ratio  & -- & -2.31 & -3.03 & -2.81 & -1.94 \\
\hline
\end{tabular}
% \vspace{-4mm}
\end{table}
\begin{table}[ht]
\centering
\scriptsize
\setlength{\tabcolsep}{5pt}
\renewcommand{\arraystretch}{1.25}
\setlength{\abovecaptionskip}{-1pt}
\caption{Three-speaker scenario (anechoic target/interference).}
\label{tab:beamforming_3spk_no_reverb}
\vspace{1mm}

\begin{tabular}{l|c|c|c|c|c}
\hline
\textbf{Metric [dB]} & \textbf{Input} & \textbf{Est. RTF} & \textbf{No RTF} & \textbf{Oracle RTF} & \textbf{LCMV} \\
\hline
\multicolumn{6}{c}{\textbf{Target speaker (enhancement)}} \\
\hline
SI-SDR  & -4.65 & 0.63 & 0.62 & 1.04 & -1.94 \\
SNR    &  1.46 & 5.74 & 6.16 & 6.02 & 2.96\\
SIR    & -3.39 & 4.90 & 5.15 & 5.49 & 6.70 \\
% Pwr Ratio & -- & 0.00 & 0.00 & 0.00 & 0.00 \\
\hline
\multicolumn{6}{c}{\textbf{Interferer 1 (suppression)}} \\
\hline
Pwr Ratio  & -- & -10.18 & -10.69 & -10.89 & -10.31 \\
\hline
\multicolumn{6}{c}{\textbf{Interferer 2 (suppression)}} \\
\hline
Pwr Ratio  & -- & -8.53 & -9.02 & -9.58 & -9.96 \\
\hline
\multicolumn{6}{c}{\textbf{Background noise (suppression)}} \\
\hline
Pwr Ratio  & -- & -4.28 & -4.69 & -4.56 & -1.50 \\
\hline
\end{tabular}
%\vspace{-6mm}
\end{table}

\begin{table}[htbp]
\centering
\scriptsize
\setlength{\tabcolsep}{6pt}
\renewcommand{\arraystretch}{1.25}
\setlength{\belowcaptionskip}{-5pt}
\setlength{\abovecaptionskip}{-1pt}
\caption{Three-speaker scenario (fully overlapped, anechoic).}
\label{tab:beamforming_3spk_no_partition}

\vspace{1mm}

\begin{tabular}{l|c|c|c}
\hline
\textbf{Metric [dB]} & \textbf{Input} & \textbf{Oracle RTF} & \textbf{No RTF} \\
\hline
\multicolumn{4}{c}{\textbf{Target speaker (enhancement)}} \\
\hline
SI-SDR & -4.65 & 1.28 & -4.62 \\
SNR    &  1.46 & 5.85 & 1.52 \\
SIR    & -3.39 & 5.74 & -3.34 \\
% SINR   & -4.65 & 2.12 & -4.60 \\
% Pwr Ratio  & -- & 0.00 & 0.00 \\
\hline
\multicolumn{4}{c}{\textbf{Interferer 1 (suppression)}} \\
\hline
% SI-SDR & -4.69 & -22.32 & -4.69 \\
Pwr Ratio  & -- & -10.91 & -0.02 \\
\hline
\multicolumn{4}{c}{\textbf{Interferer 2 (suppression)}} \\
\hline
% SI-SDR & -4.67 & -19.04 & -4.69 \\
Pwr Ratio  & -- & -9.81 & -0.04 \\
\hline
\multicolumn{4}{c}{\textbf{Background noise (suppression)}} \\
\hline
Pwr Ratio  & -- & -4.39 & -0.05 \\
\hline
\end{tabular}
\end{table}

\begin{figure}[h!]
\vspace{-5pt}
\centering
\begin{subfigure}{0.45\linewidth}
    \includegraphics[width=\linewidth]{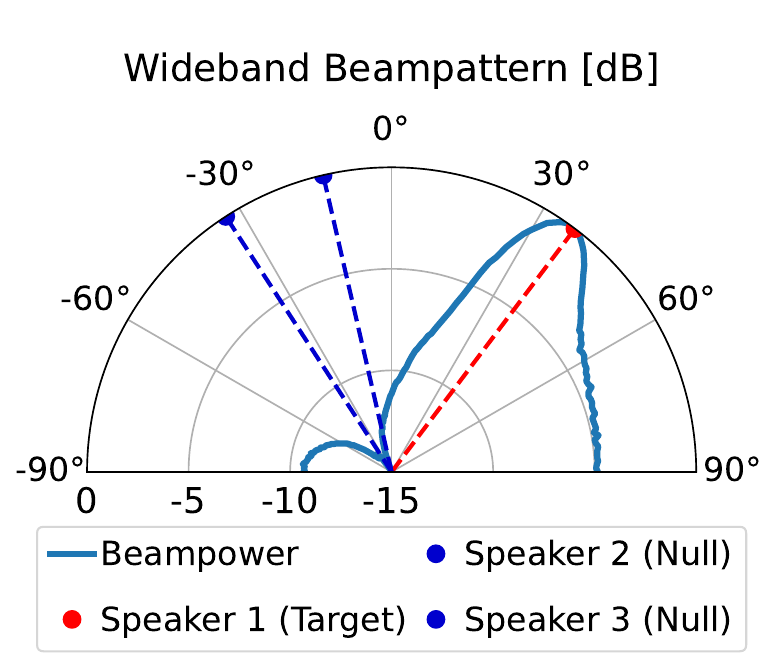}
    \caption{Estimated RTF}
\end{subfigure}
\hfill
\begin{subfigure}{0.45\linewidth}
    \includegraphics[width=\linewidth]{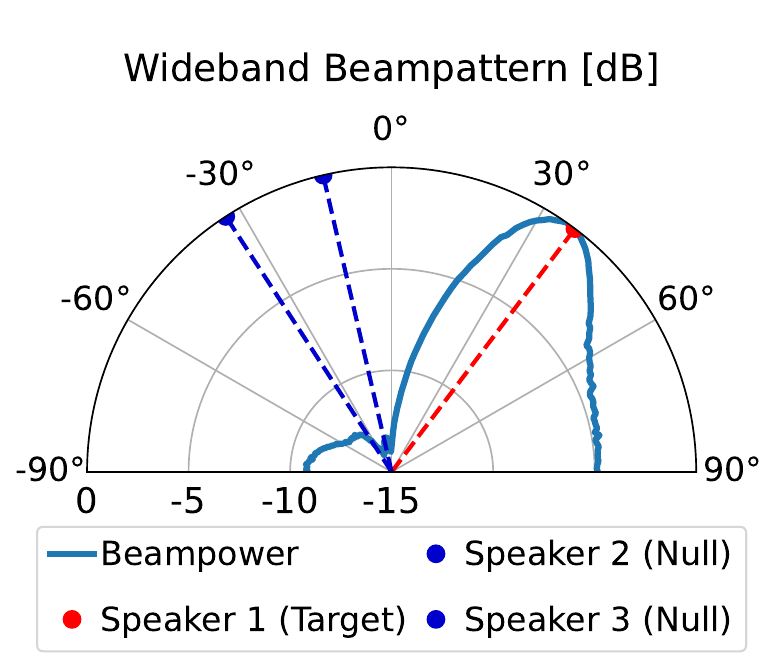}
    \caption{Oracle RTF}
\end{subfigure}
\vspace{2mm}

\begin{subfigure}{0.45\linewidth}
    \includegraphics[width=\linewidth]{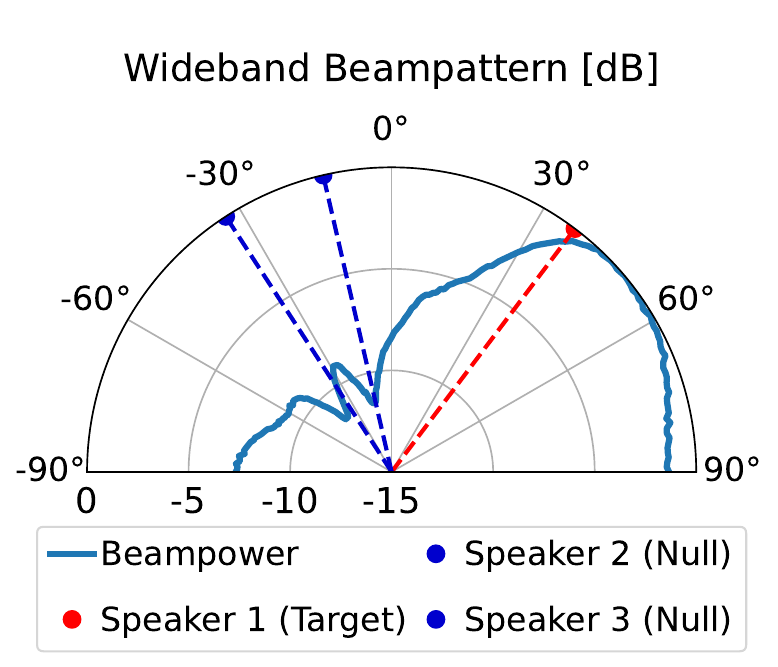}
    \caption{No RTF}
\end{subfigure}
\hfill
\begin{subfigure}{0.45\linewidth}
    \includegraphics[width=\linewidth]{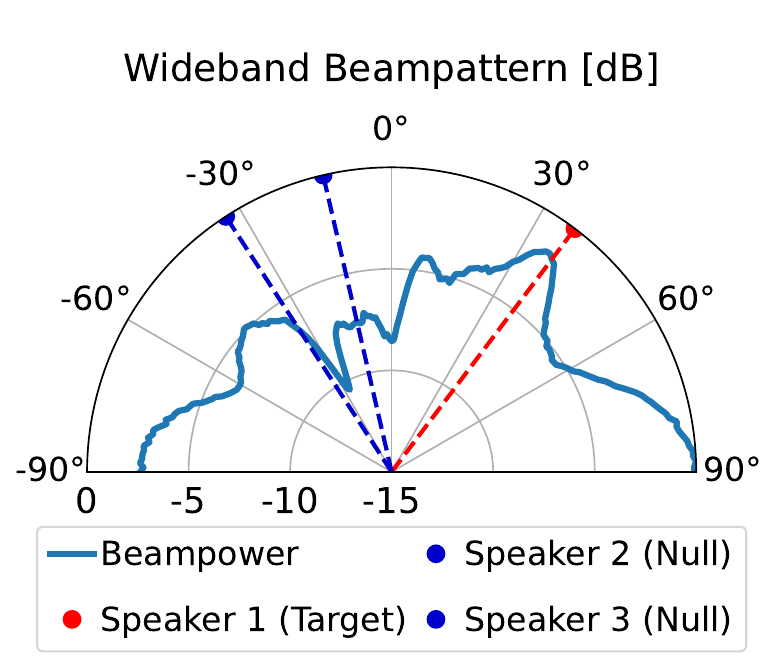}
    \caption{LCMV}
\end{subfigure}
\setlength{\belowcaptionskip}{-15pt}
\setlength{\abovecaptionskip}{0pt}
\caption{Beampattern for different methods in anechoic case.}
%\vspace{-4mm}
\label{fig:beampattern_collage_3_speakers}
\end{figure}

\section{Conclusions}
\label{sec:Conclusions}

In this work, we propose a fully DNN-based beamforming framework for target-speaker enhancement in multi-speaker environments that leverages explicit spatial guidance. The proposed method combines RTF-based guidance with an adaptive loss
% inspired by constrained optimization,
with progressively increasing constraint weights,
enabling the network to preserve the target speaker and suppress interfering speakers within a fully learned beamforming framework.
% The results demonstrate that the proposed approach learns spatially selective filtering behavior, producing focused beampatterns and effective interference suppression that outperform those of a classical LCMV beamformer constructed
% from the estimated spatial signatures in several evaluated scenarios.
The results demonstrate that the proposed approach learns spatially selective filtering, producing focused beampatterns and achieving a better overall balance of target enhancement, interference suppression, and noise attenuation than an LCMV beamformer constructed from the estimated spatial signatures.
Moreover, the proposed method exhibits lower sidelobe levels, resulting in more focused and spatially selective beampatterns. Overall, the proposed framework highlights the potential of incorporating explicit spatial constraints and priors into interpretable and robust DNN-based multichannel speech enhancement systems.

\clearpage
% \balance
{
\bibliographystyle{IEEEbib}
\balance
\bibliography{DNN_BASED_BEAMFORMER}
}
\end{document}

%% file: figures/proposed_model.tex
\tikzset{
  >=Latex,
  block/.style={draw, rounded corners=3pt, thick, align=center,
                minimum width=20mm, minimum height=10mm, blur shadow},
  blockISTFT/.style={draw, rounded corners=3pt, thick, align=center,
                minimum width=20mm, minimum height=10mm},                
  tinyblock/.style={draw, rounded corners=2pt, thick, align=center,
                    minimum width=12mm, minimum height=8mm, fill=specbg},
  tinyblockATTN/.style={draw, rounded corners=2pt, thick, align=center,
                    minimum width=13mm, minimum height=16mm, fill=specbg, blur shadow},
  arrow/.style={-Latex, very thick},
  stream/.style={draw, thick, minimum width=8mm, minimum height=18mm,
                 rounded corners=2pt, fill=white},
  mult/.style={
    draw, circle, thick, fill=white, inner sep=0pt,
    minimum size=6mm,     % circle size; tweak 7–10mm to taste
    path picture={
      % draw the X inside the circle
      \draw[line width=0.9pt]
        ($(path picture bounding box.south west)+(1.0pt,1.0pt)$) --
        ($(path picture bounding box.north east)+(-1.0pt,-1.0pt)$);
      \draw[line width=0.9pt]
        ($(path picture bounding box.north west)+(1.0pt,-1.0pt)$) --
        ($(path picture bounding box.south east)+(-1.0pt,1.0pt)$);
    },
  },
waveglyph/.style={},
  pic wave/.style n args={3}{ % width, height, cycles
    code={
      \def\W{#1}
      \def\H{#2}
      \def\C{#3}
      \draw[line width=0.6pt]
        plot[domain=0:\W, samples=60]
        (\x,{ \H*sin(360*\C*\x/\W) });
    }
  },
  wavstack/.style={
    thick, rounded corners=2pt, fill=white,
    minimum width=8mm, minimum height=18mm
  },
  pics/wave/.style n args={3}{
  code={
    \def\W{#1}
    \def\H{#2}
    \def\C{#3}
% amplitude function (keep exactly your 6 elements)
\def\amp{
  (
    \H * (
         0.25
       + 0.85*exp(-18*((\x+0.5)-0.25)^2)
       + 0.65*exp(-22*((\x+0.5)-0.55)^2)
       + 0.45*exp(-26*((\x+0.5)-0.82)^2)
    )
    * abs(
        0.90*sin(360*\C*(10*(\x+0.5)))
      + 0.35*sin(360*\C*(20*(\x+0.5)) + 20)
      + 0.18*sin(360*\C*(30*(\x+0.5)) - 55)
    )
  )
}

% upper half bars
\draw[line width=0.9pt, line cap=round]
  plot[domain=-0.5:0.5, samples=19, variable=\x, ycomb]
  ({\x*\W},{\amp});

% lower half bars (mirror)
\draw[line width=0.9pt, line cap=round]
  plot[domain=-0.5:0.5, samples=19, variable=\x, ycomb]
  ({\x*\W},{-\amp});
  }
},
}
\newcommand{\bigvdots}{%
\begin{tikzpicture}[baseline=-0.6ex]
  \fill (0,0) circle (1.2pt);
  \fill (0,-0.18) circle (1.2pt);
  \fill (0,-0.36) circle (1.2pt);
\end{tikzpicture}%
}

% \begin{tikzpicture}[
%   font=\fontfamily{qag}\selectfont\small,
%   every node/.append style={font=\fontfamily{qag}\selectfont\small}
% ]
% \begin{tikzpicture}[
%   font=\fontfamily{qhv}\selectfont\small,
%   every node/.append style={font=\fontfamily{qhv}\selectfont\small}
% ]
\begin{tikzpicture}[
  font=\small,
  every node/.style={font=\small}
]
% ---------- Input (8 channels) ----------
% \node[stream, label={[align=center]above: \large 8-ch noisy\\ \large inputs}] (in) {};
% \foreach \y in {1,...,8} {
%   \draw ($(in.center)+(-3mm,{-0.9cm+0.20cm*\y})$) -- ++(6mm,0);
% }
\node[
  wavstack,
  label={[align=center, xshift=-3mm, yshift=2mm]above:
    \large Multichannel noisy\\ \large inputs}
] (in) {};
% \foreach \k in {0,...,7} {
%   \pic at ($(in.north)+(0mm,{-3.0mm-1.7mm*\k})$) {wave={10.5mm}{2.9mm}{1.2}};
% }
% top wave
\pic at ($(in.north)+(-3mm,-2mm)$) {wave={14.5mm}{3.2mm}{1.2}};

% dots in the middle
\node at ($(in.center)+(-3mm,0mm)$) {\bigvdots};
\node[left=6mm of in] (spk) {
    \includegraphics[height=1.5cm]{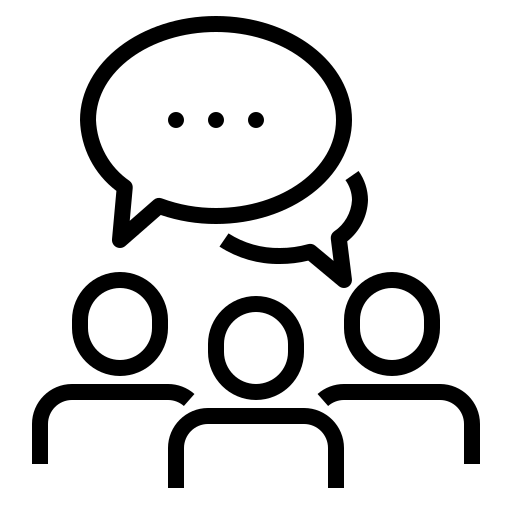}
};
% bottom wave
\pic at ($(in.south)+(-2.5mm,2mm)$) {wave={14.5mm}{3.2mm}{1.2}};
% ---------- Shared STFT ----------
\node[block, fill=stftcol, right=10mm of in] (stft) {\Large STFT};

% split point
\coordinate (split) at ($(stft.east)+(14mm,0)$);

\draw[arrow] (in) -- (stft);
% \draw[arrow] (stft.east) -- (split);
\draw[very thick] (stft.east) -- node[above,yshift=1pt] {\LARGE $\mathbf y$} (split);

% ===================== Left branch =====================
% \node[block, fill=leftcol!60, right=22mm of stft, yshift=0mm] (rtfL) {\large RTF\\ \large Estimation};
% --- RTF estimation shown as 3 stacked cards ---
\coordinate (rtfpos) at ($(stft.east)+(30mm,0)$);

% back card
\node[
  draw=black!70, rounded corners=2pt, thick, fill=teal!10,
  minimum width=22mm, minimum height=15mm
] (rtfLback) at ($(rtfpos)+(2.4mm,2.4mm)$) {};

% middle card
\node[
  draw=black!70, rounded corners=2pt, thick, fill=teal!20,
  minimum width=22mm, minimum height=15mm
] (rtfLmid) at ($(rtfpos)+(1.2mm,1.2mm)$) {};

% front card
\node[
  draw=black!70, rounded corners=2pt, thick, fill=teal!35, 
  minimum width=22mm, minimum height=15mm,
  align=center
] (rtfL) at (rtfpos) {\large RTF\\ \large Estimation};

% NEW: small Attention block
\node[tinyblockATTN, fill=orange!15!yellow!10, right=13mm of rtfL] (attnL) {\large Attn.};
% arrows from each RTF card to attention block
% \draw[arrow] (rtfLback.east) -- ($(attnL.west)+(0,2.5mm)$);
% \draw[arrow]  ($(rtfL.east)+(1.2mm,0)$) -- (attnL);    
% \draw[arrow]  ($(rtfL.east)+(0mm,-2.5mm)$)     -- ($(attnL.west)+(0,-2.5mm)$);
% U-Net after attention
\node[block, fill=leftcol!80, right=10mm of attnL] (unetL){\large U\text{-}Net};

\node[mult, right=15mm of unetL] (mulL) { };
\node[tinyblock, right=10mm of mulL] (specL) {\LARGE $\hat{s}$};
\node[blockISTFT, fill=istftcol, right=10mm of specL] (istftL) {\large ISTFT};

% Output boxes: define a shared x-position so they align perfectly
\coordinate (outX) at ($(istftL.east)+(18mm,0)$);

% \node[draw, rounded corners=2pt, minimum width=16mm, minimum height=10mm, align=center]
%   (outL) at (outX |- istftL.center) {\large Left\\ \large out};

% connections (Left)
\draw[arrow] (split) -- (rtfL.west);          % y -> RTF
           % RTF -> Attn
\draw[arrow] (attnL) -- (unetL);              % Attn -> U-Net

% y -> Attn (fusion input)
% \draw[arrow] (split) |- ($(attnL.west)+(-2mm,0)$) -- (attnL.west);

% keep your top-down arrow into U-Net (optional)
\draw[arrow] (split) |- ++(0,17mm) -| (attnL.north);

\draw[arrow] (unetL) -- node[above] {\LARGE $\mathbf w$} (mulL);

% BYPASS: y to multiplier (upper route)
\draw[arrow] (split) |- ++(0,17mm) -| (mulL.north);

\draw[arrow] (mulL) -- (specL);
\draw[arrow] (specL) -- (istftL);
% \draw[arrow] (istftL) -- (outL);
\draw[arrow] (rtfLback.east) -- ($(attnL.west)+(0,4mm)$);
\draw[arrow] (rtfLmid.east)  -- ($(attnL.west)+(0,0mm)$);
\draw[arrow] (rtfL.east)     -- ($(attnL.west)+(0,-4mm)$);

% ---------- Branch headers ----------

\end{tikzpicture}